\documentstyle[12pt]{article}
\def\PRL #1 #2 #3{{\sl Phys. Rev. Lett.} {\bf#1} (#2) #3}
\def\NPB #1 #2 #3{{\sl Nucl. Phys.} {\bf B #1} (#2) #3}
\def\NPBFS #1 #2 #3 #4{{\sl Nucl. Phys.} {\bf B #2} [FS#1] (#3) #4}
\def\CMP #1 #2 #3{{\sl Commun. Math. Phys.} {\bf #1} (#2) #3}
\def\PRD #1 #2 #3{{\sl Phys. Rev.} {\bf D #1} (#2) #3}
\def\PLA #1 #2 #3{{\sl Phys. Lett.} {\bf #1A} (#2) #3}
\def\PLB #1 #2 #3{{\sl Phys. Lett.} {\bf B #1} (#2) #3}
\def\JMP #1 #2 #3{{\sl J. Math. Phys.} {\bf #1} (#2) #3}
\def\PTP #1 #2 #3{{\sl Prog. Theor. Phys.} {\bf #1} (#2) #3}
\def\SPTP #1 #2 #3{{\sl Suppl. Prog. Theor. Phys.} {\bf #1} (#2) #3}
\def\AoP #1 #2 #3{{\sl Ann. of Phys.} {\bf #1} (#2) #3}
\def\PNAS #1 #2 #3{{\sl Proc. Natl. Acad. Sci. USA} {\bf #1} (#2) #3}
\def\RMP #1 #2 #3{{\sl Rev. Mod. Phys.} {\bf #1} (#2) #3}
\def\PR #1 #2 #3{{\sl Phys. Reports} {\bf #1} (#2) #3}
\def\AoM #1 #2 #3{{\sl Ann. of Math.} {\bf #1} (#2) #3}
\def\UMN #1 #2 #3{{\sl Usp. Mat. Nauk} {\bf #1} (#2) #3}
\def\FAP #1 #2 #3{{\sl Funkt. Anal. Prilozheniya} {\bf #1} (#2) #3}
\def\FAaIA #1 #2 #3{{\sl Functional Analysis and Its Application} {\bf
#1} (#2) #3}
\def\BAMS #1 #2 #3{{\sl Bull. Am. Math. Soc.} {\bf #1} (#2)
#3} \def\TAMS #1 #2 #3{{\sl Trans. Am. Math. Soc.} {\bf #1} (#2) #3}
\def\InvM #1 #2 #3{{\sl Invent. Math.} {\bf #1} (#2) #3}
\def\LMP #1 #2 #3{{\sl Letters in Math. Phys.} {\bf #1} (#2) #3}
\def\IJMPA #1 #2 #3{{\sl Int. J. Mod. Phys.} {\bf A #1} (#2) #3}
\def\AdM #1 #2 #3{{\sl Advances in Math.} {\bf #1} (#2) #3}
\def\RMaP #1 #2 #3{{\sl Reports on Math. Phys.} {\bf #1} (#2) #3}
\def\IJM #1 #2 #3{{\sl Ill. J. Math.} {\bf #1} (#2) #3}
\def\APP #1 #2 #3{{\sl Acta Phys. Polon.} {\bf #1} (#2) #3}
\def\TMP #1 #2 #3{{\sl Theor. Mat. Phys.} {\bf #1} (#2) #3}
\def\JPA #1 #2 #3{{\sl J. Physics} {\bf A#1} (#2) #3}
\def\JSM #1 #2 #3{{\sl J. Soviet Math.} {\bf #1} (#2) #3}
\def\MPLA #1 #2 #3{{\sl Mod. Phys. Lett.} {\bf A #1} (#2) #3}
\def\JETP #1 #2 #3{{\sl Sov. Phys. JETP} {\bf #1} (#2) #3}
\def\JETPL #1 #2 #3{{\sl  Sov. Phys. JETP Lett.} {\bf #1} (#2) #3}
\def\PHSA #1 #2 #3{{\sl Physica} {\bf A #1} (#2) #3}
\def\CQG #1 #2 #3{{\sl Class. Quantum Grav.} {\bf #1} (#2) #3}
\def\SJNP #1 #2 #3{{\sl Sov. J. Nucl. Phys. (Yadern.Fiz.)} {\bf #1} (#2) #3}
\def\a{\alpha}\def\b{\beta}\def\g{\gamma}\def\d{\delta}\def\e{\epsilon}

\def\Th{\Theta}\def\Om{\Omega}\def\G{\Gamma}

\newcommand{\p}[1]{(\ref{#1})}
\begin{document}
\renewcommand{\thefootnote}{\fnsymbol{footnote}}
\begin{center}
{\bf
GENERALIZED ACTION PRINCIPLE and GEOMETRIC APPROACH
for SUPERSTRINGS and SUPER--$P$--BRANES}

\renewcommand{\thefootnote}{\dagger} \vspace{0.2cm}

{\bf Igor A. Bandos,}
\footnote{e-mail:  kfti@rocket.kharkov.ua  ~~~
\\ Work was
supported  in  part  by  the
INTAS and Dutch
Government grant {\bf 94-- 2317}.}

\vspace{0.2cm}  {\it
{\small Institute  for  Theoretical  Physics,  NSC  KFTI } {\it
310108, Kharkov, Ukraine}} \\

\setcounter{page}1
\renewcommand{\thefootnote}{\arabic{footnote}}
\setcounter{footnote}0

\vspace{0.2cm}
\end{center}

{\small
\begin{quotation}
Most of results described in this talk were obtained in collaboration
with Dmitrij P. Sorokin and our teacher Dmitrij V. Volkov who, to our
great sorrow,  untimely leaved us this January.
\end{quotation} }

{\small
In this short printed version of my talk I  describe
the generalized action for super--$p$-- branes
[1] which can be used
to construct the doubly supersymmetric generalization
[2] of the geometric approach
[3], and  discuss one of the direction
of its possible application
[4].

{\bf 1.}
The generalized action
for super--$p$-- brane
[1]
\begin{equation}\label{1}
S_{D,p}=
\int\limits^{}_{{\cal M}^{p+1}}  {\cal L}_{p+1} =
\int\limits^{}_{{\cal M}_{p+1}}
\left(
-{{(-1)^p}\over{p!}} (E^{a}e^{a_1}...e^{a_p}  -
{p\over{(p+1)}}  e^{a} e^{a_1}... e^{a_{p}})
\varepsilon_{aa_1...a_{p}} + {\cal L}_{p+1}^{WZ} \right)
\end{equation}
is the integral of $(p+1)$-- form ${\cal L}_{p+1}$ over
arbitrary $(p+1)$ -- dimensional bosonic surface
${\cal M}^{p+1} = \{ (\xi^m , \eta^{\mu q} ):  \eta^{\mu q}=
\eta^{\mu q}(\xi)\} $ in the world volume superspace
$ \Sigma^{(p+1|n)} = \{ (\xi^m , \eta^{\mu q} )\}$ of the super-- $p$--
brane $(m=0,1,...,p)$.  Lagrangian two-- form \p{1} contains the Wess--
Zumino term ${\cal L}^{WZ}_{p+1}$
($~d{\cal
L}^{WZ}_{p+1} = - i \Pi^{\underline{m}_{p+1}}\ldots
\Pi^{\underline{m}_1} d\Theta \Gamma_{\underline{m}_1\ldots
\underline{m}_p} d\Theta $)
and is constructed from some
of the basic one forms of target superspace
\begin{equation}\label{3}
\Pi^{\underline{m}}= dX^{\underline{m}} - i d\Theta
\Gamma_{\underline{m}} \Theta, \qquad   d\Theta^{\underline{\mu}}
\qquad \underline{m} = 0, \ldots , (D-1)
\end{equation}
\begin{equation}\label{4}
E^{\underline{a}} = (E^a, E^i)
= \Pi^{\underline{m}} u_{\underline{m}}^{\underline{a}} =
 ( \Pi^{\underline{m}} u_{\underline{m}}^{a},
\Pi^{\underline{m}} u_{\underline{m}}^{i})
\qquad {\underline{a} = 0, \ldots , (D-1)
; ~~ a = 0, \ldots , p}
\end{equation}
\begin{equation}\label{5}
E^{\underline{\a}} = (E^{\a q}, E^{\a \dot{q}})
= d\Theta^{\underline{\mu}} v_{\underline{\mu}}^{\underline{\a}} =
 (d\Theta^{\underline{\mu}} v_{\underline{\mu}}^{\a q},
d\Theta^{\underline{\mu}} v_{\underline{\mu}}^{\a \dot{q}}),
\end{equation}
and world volume superspace
$ e^A = (e^a, e^{\a q}) = d\xi^m e_m^A + d\eta^{\mu q} e_{\mu q}^A $
using the external product of the forms only.
Supervielbein  of flat target superspace
$E^{\underline{A}} =(E^{\underline{a}}, E^{\underline{\a }})~$ \p{4},
\p{5} differs from the standard one \p{3} by Lorentz rotation, which
vector and spinor representations are given by the matrices
$u^{~\underline{a}}_{\underline{m}}$ and
$v^{\underline{\a}}_{\underline{\mu}}$
$$||u^{~\underline{a}}_{\underline{m}}|| =
||(u^{a}_{\underline{m}}, u^{~i}_{\underline{m}})|| ~~ \in ~~ SO(1,D-1)
~~~~\Leftrightarrow ~~~~
u^{~\underline{a}}_{\underline{m}} \eta^{\underline{m}\underline{n}}
u^{~\underline{b}}_{\underline{n}} = \eta^{\underline{a}\underline{b}}
$$ $$
||v^{~\underline{\a}}_{\underline{\mu}}|| =
||(v^{\a q}_{\underline{\mu}}, u^{\a \dot{q}}_{\underline{\mu}})||
~~ \in ~~ Spin(1,D-1) $$
  Hence, $u$ and $v$ matrices
(vector and spinor Lorentz harmonics, see
[5,6,2] and refs. in [2])
are related by the conditions of the conservation of the
$D$-- dimensional $\g$-- matrices
$$ u^{~\underline{a}}_{\underline{m}}
\Gamma^{\underline{m}}_{\underline{\mu}\underline{\nu}}
= v_{\underline{\mu}}^{~\underline{\a}}
\Gamma^{\underline{a}}_{\underline{\a }\underline{\b }} v^{~\underline{\b}}
_{\underline{\mu}},
\qquad
u^{~\underline{a}}_{\underline{m}}
\Gamma_{\underline{a}}^{\underline{\a }\underline{\b }}
= v_{\underline{\mu}}^{~\underline{\a}}
\Gamma_{\underline{m}}^{\underline{\mu }\underline{\nu }}
v^{~\underline{\b}} _{\underline{\mu}},
$$
Their differentials
$ d u^{~\underline{a}}_{\underline{m}} =
 u^{~\underline{b}}_{\underline{m}}
 \Om^{~\underline{a}}_{\underline{b}} (d), ~~~
 d v^{~\underline{\a}}_{\underline{\mu}} \propto
 1/4 v^{~\underline{\b}}_{\underline{\mu}}
 (\Gamma_{\underline{a}\underline{b}})^{~\underline{\a}}_{\underline{\b}}
 \Om^{~\underline{a}\underline{b}} (d) , ~~
$
are expressed in terms of the
$so(1,D-1)$ valued Cartan $1$-- form
\begin{equation}\label{9}
\Om^{\underline{a}\underline{b}} = -
\Om^{\underline{b}\underline{a}} =
\left(
\matrix{ \Om^{ab} & \Om^{aj} \cr
       - \Om^{bi} & \Om^{ij} \cr}
        \right)
= u^{~\underline{a}}_{\underline{m}} d u^{\underline{b}\underline{m}}
\propto
dv^{~\underline{\a}}_{\underline{\mu}}
(\Gamma^{~\underline{a}\underline{b}})_{\underline{\a }}^{\underline{\b}}
v^{~\underline{\mu }}_{\underline{\b}}
\end{equation}

In  the functional \p{1} all the variables shall be considered as
world volume superfields
\\$
X^{\underline{m}} = X^{\underline{m}} (\xi , \eta ) , ~~~
\Theta^{\underline{\mu}} = \Theta^{\underline{\mu}}  (\xi , \eta ) , ~~~
u^{~\underline{a}}_{\underline{m}} =
u^{~\underline{a}}_{\underline{m}} (\xi , \eta ) , ~~~
e^{~\underline{A}}_{\underline{M}} =
e^{~\underline{A}}_{\underline{M}}
(\xi , \eta ) , ~~~...
~~$
but taken on the surface ${\cal M}^{(p+1)}: ~~ \eta = \eta (\xi ) , 
~~~
X^{\underline{m}} = X^{\underline{m}} (\xi , \eta (\xi)) , ~~~ ...
e^{~\underline{A}}_{\underline{M}} =
e^{~\underline{A}}_{\underline{M}}
(\xi , \eta (\xi ) ) , ~~~...
$

The generalized action concept
consists in the requirement that the variation of the functional \p{1}
should vanishes for arbitrary variations of the (super)fields involved
 as well as for arbitrary variations of the surface ${\cal M}^{(p+1)}$.
 For the Lagrangian form under consideration it can be proved (see 
[1] and Refs. terein) that the variation with respect to the surface ${\cal
M}^{(p+1)}$ (i.e. $\d S / d \eta (\xi) = 0 $) does not lead to new
 equations of  motion. However, the arbitrariness of  surface ${\cal
 M}^{(p+1)}$ gives the possibility to consider another equations of motion
 ($\d S / \d \Theta = 0$, etc. ) as superfield equations, i.e.  as
equations for the forms and superfields defined on the whole world volume
superspace $\Sigma^{(p+1|n)}$.

The detailed consideration of the properties of the generalized action
\p{1} can be found in Refs. [1].

{\bf 2.} A pure  bosonic limit
[6] $S^0_{D,p}$ of the generalized action \p{1} 
is provided by
the substitution
 ${\cal M}^{(p+1)} \rightarrow {\cal M}^{(p+1)} =
\{ (\xi^m , \eta^{\mu p} ) : \eta^{\mu p} = 0 \} $,  $\Theta = 0$

Its equations of motion split naturally into rheotropic conditions
[1]
\begin{equation}\label{11}
E^a  = e^a , \qquad
E^i = 0, \qquad \Rightarrow \qquad
dX^{\underline{m}} = e^a u^{\underline{m}}_a
\end{equation}
and proper dynamical equation
$u^{i\underline{m}}\d S /\d X^{\underline{m}} = 0 $, which can be written in terms of the
pull-- back $\Om^{ai}= d\xi^m \Om_m^{ai}$ of the covariant Cartan form
$\Om^{ai} = u^a_{\underline{m}} du^{i\underline{m}}$ \p{9}
\begin{equation}\label{12}
\Om^{ai}(\nabla_a)\equiv e_a^m \Om_m^{ai} = 0 ,
\end{equation}

Passing from Eqs. \p{11} to
 their selfconsistency (integrability) conditions
($ddX= 0 = d(e^a u^{\underline{m}}_{\underline{a}})$) we can exclude the
embedding functions $X(\xi)$ as well as harmonic fields $u(\xi)$ 
($v(\xi)$) from the consideration and get the equations 
\begin{equation}\label{13}
e_a \Om^{ai} = 0 , \qquad
T^a \equiv {\cal D} e^{a} \equiv
de^a - e_b \Om^{ba} = 0 ,
\end{equation}
 written in terms of intrinsic vielbeins $e^a = d\xi^m e^a_m$ and
Cartan forms \p{9} only. The later satisfy the Maurer-- Cartan equations
\begin{equation}\label{15}
d\Omega ^{\underline a\underline b} -
\Omega ^{\underline a}_{~\underline c} \Omega ^{\underline c \underline b}
= 0,  \qquad
 \cases {
{\cal D}\Omega ^{ai} \equiv
d\Omega ^{ai} -
\Omega ^{a}_{~b} \Omega ^{bi} +
\Omega ^{aj} \Omega ^{ji} = 0,  \cr
{R}^{ab}(d,d)=
d\Omega ^{ab} -
\Omega ^{a}_{~c} \Omega ^{cb} =
\Omega ^{ai} \Omega^{bi},  \cr
{R}^{ij}(d,d)=
d\Omega ^{ij} +
\Omega ^{ij^\prime} \Omega^{j^\prime j} = -
\Omega ^{ai} \Omega^{~j}_{a},  \cr }
\end{equation}

The equations \p{12} -- \p{15} describe the minimal embedding of
the $p$-- brane world volume into the flat target space-- time and are
referred as geometric approach equations
[3,2,7].

{\bf 3. Some properties of doubly supersymmetric geometric approach}.
Independent equations of motion for generalized action \p{1} are the
rheotropic conditions
\begin{equation}\label{16}
E^a = \Pi^{\underline{m}} u_{\underline{m}}^{a} = e^a, \qquad
E^i =  \Pi^{\underline{m}} u_{\underline{m}}^{i} = 0 ,  \qquad
D_{\a q} \Theta^{\underline{\mu}} v_{\underline{\mu}\a \dot{q}} = 0
\end{equation}
and proper dynamical equation
$(\g^a)_\b ^\a D_{a} \Theta^{\underline{\mu}} v_{\underline{\mu}\a 
\dot {q}} =
0$
It is important that Eq. \p{16}
$\Pi^{\underline{m}} = d X^{\underline{m}} - i d\Th
\G^{\underline{m}} \Th = e^a$
contains, in particular, Geometrodynamical condition [8]
$
\Pi^{\underline{m}}_{\a q} = D_{\a q} X^{\underline{m}} - i
D_{\a q} \Th
\G^{\underline{m}} \Th = 0
$,
which was a starting point for previous constructions of world sheet
superfield formulations of superparticles and superstrings
[8,2], as well as the twistor--
like solution
$
\Pi^{\underline{m}}_{a} = D_{a } X^{\underline{m}} - i
D_{a } \Th
\G^{\underline{m}} \Th = u_a^{\underline{m}}
~$
of the Virasoro-- like constraint $\Pi^{\underline{m}}_a
\Pi^{\underline{m}}_b = \eta_{ab} $.

The geometric approach equations for super-- $p$-- branes
[2,1,9]
can be obtained
by considering  the integrability conditions for Eqs. \p{16}
and for the relations
\begin{equation}\label{19}
E^{ \a q} \equiv
 d\Th^{\underline \mu}v_{\underline \mu}^{ \a q} = e^{\a q}, \qquad
E^{ \a \dot{q}} \equiv
 d\Th^{\underline \mu}v_{\underline \mu}^{ \a \dot{q}} =
 e^a \psi_{a \a \dot{q}}, \qquad
\end{equation}
 which collect the third equation from \p{16} and conventional rheotropic
 conditions [1,9].
 They are
formulated in terms of the world volume vielbein $e^A = (e^a, e^{\a q})$
and independent components of the pull-- backs of the Cartan forms \p{9},
one of which coincides with the superfield 
$\psi_a^{\a \dot{q}} = D_a
\Theta^{\underline{\mu}} v_{\underline{\mu}}^{\a \dot{q}}$ \p{19}
[2,9].

It shall be stressed, that
{\it the world volume supergravity  constraints follows from
the equations of motion for the generalized action \p{1}}.
Indeed, the integrability conditions for the first of Eqs. \p{16} acquire
the form
$T^a=De^a\equiv de^a-\Om^a_be^b= -2ie^{\a q} e^{\b p} \g^a_{\a\b}
$ after some algebra. This equation contains the most essential
supergravity constraint $T^a_{\a p\b q}=-2i\g^a_{\a\b}\d_{pq}$. Other
constraints of world volume supergravity are conventional and also appear
as  selfconsistency conditions, because we us the freedom of redefinition
of the geometrical quantities, which are absent in the action, to fix them
to be induced by the embedding.

{\bf 4.}
In conclusion, let us consider
bosonic  $p$-- brane interacting with generalized Kalb-- Ramond
field
$B_{p+1} = dX^{\underline{m}_{p+1}}  ... dX^{\underline{m}_{1}}
B_{\underline{m}_{1}... \underline{m}_{p+1}} (X)
$. 
The action functional $S = S^{0}_{D,p} + S^{int}_{D,p} $
is the sum of the free $p$-- brane action [6] 
(see item 2) and
the interaction term
 $
S^{int}_{D,p} = - \int_{{\cal{M}}^{p+1}}  B_{p+1}
$.
It can be proved
[4], that the geometric approach equations \p{13}-- \p{15}
(which describes world volume as a surface embedded into space-- time) remains the same for the system under consideration,
and Eq. \p{12} is replaced by
$$
\Om^{ai}(\nabla_a) = e_a \Om^{ai} = {1 \over {(p+1)!}} 
\e _{a_0 ... a_p }
u^{a_0 \underline{m}_0}\ldots u^{a_p\underline{m}_p} u^{i\underline{m}_{p+1}} 
H_{\underline{m}_0 \ldots \underline{m}_{p+1}} (X(\xi)) ,
$$
where
$
H_{\underline{m}_0 \ldots \underline{m}_{p+1}} (X(\xi)) =
(p+1) \partial_{[\underline{m}_0}
B_{\underline{m}_1 \ldots \underline{m}_{p+1}]} (X(\xi))
$.

Hence, the world volume is embedded as a nonminimal surface and its main curvature
is defined by the field strength of the generalized Kalb-- Ramond field,
{\it which can be considered as arbitrary function of the coordinates $X(\xi)$}.

The general theorems about local isometric embedding
[10]  guarantees that, if the dimension of target space time
is $D \geq (p+1)(p+2)/2$, then we can describe arbitrary curved
 $d=(p+1)$-- dimensional surface in such a way.

Hence, the considered model can be regarded as a model for description of
$d=(p+1)$ -- dimensional gravity providing the dynamical ground for the
embedding approach used before for investigation of General Relativity
[10].
This realizes the idea of Regge and Teitelboim
[11] about string-- like description of
Gravity.

The model for $d=4$ supergravity is provided by $D=10$ $3$-- brane
\footnote{It is interesting that the number $D=10$ of space time dimensions is distinguished by
superstring theory, and that a $3$-- brane supersymmetric soliton exists
in $D=10$ type $IIB$ superstring theory
[12]. Moreover, this soliton is exceptional in some reasons
[12].}.

 The generalized action principle provides a ground for supersymmetric
 generalization of such construction.

{\small
\begin{center}
{\bf \large References}
\end{center}
1. I. Bandos, D. Sorokin and D. Volkov,
\PLB 352 1995 269.
D.V. Volkov, {\sl Generalized Action Principle for superstrings and
supermembrane,} {\bf
hep-th/9512103}, In: {\sl Proc. Intern. Conference ''SUSY 95''. Paris,
France.}
\\ 2. I. Bandos, P. Pasti, D. Sorokin, M. Tonin and D. Volkov,
{\sl Nucl.Phys.}{\bf B446} (1995) 79.
\\ 3. B. M.  Barbashov and V. V. Nesterenko, {\sl Introduction to the
relativistic string theory} World Scientific, 1990; {\sl and refs. therein}
\\ 4. I. Bandos and W. Kummer,
 {\sl  P-branes in D--dimensional Kalb--Ramond Background as (p+1)--
 dimensional Gravity} (in preparation), 
I. Bandos, {\sl String--Like Description of Gravity and Possible Applications 
for $F$--theory}, {\sl Preprint} {\bf TUW 96-16}, {\bf hep--th/9608093}.  
\\ 5. E. Sokatchev, {\sl Phys. Lett.} {\bf B169} (1987) 209; \CQG 4 1987
237.
\\ 6. I. A. Bandos and A. A. Zheltukhin,
 {\sl Fortschr. Phys.} {\bf 41} (1993) 619.
{\sl Phys. Part. Nucl.} {\bf 25} (1994) 1065.
{\sl Class. Quantum Grav.} {\bf 12} (1995) 609
{\it and refs. therein}
\\ 7.I.Bandos,
{\bf hep--th/9510216}, submitted to {\sl Phys.Lett.}{\bf B}.
\\ 8. D. Sorokin, V. Tkach and D. V. Volkov, {\sl Mod. Phys. Lett.} {\bf
A4} (1989) 901; F. Delduc, A. Galperin, P. Howe and E. Sokatchev, {\sl
Phys. Rev.} {\bf D47} (1992) 587, and refs. in
[2].
\\ 9. I. Bandos, D. Sorokin and D. Volkov,
{\sl Phys.Lett.} {\bf B372} (1996) 77,
I. Bandos,
{\sl Doubly supersymmetric geometric approach for heterotic
string:  From Generalized Action Principle to exactly solvable nonlinear
equations}
{\bf hep-th/9510213},
{\sl of D.P.
Sorokin on this Conference and refs. therein}.
 \\ 10. L.P. Eisenhart, {\it
 Riemannian  Geometry}, Princeton Univ. Press. 1926
 {\it and refs. therein}.
 D. Kramer, H. Stephani, M. Mac-- Callum, E. Herlt, Ed. E. Schmutzer, {\sl
 Exact Solutions of Einstein Field Equation}, Berlin, 1980,
(and references therein).
\\ 11. T. Regge and C. Teitelboim, {\sl Proc. First Marcel Grossman
Meeting, Trieste, Italy, 1975. } ed. R. Ruffini (Amsterdam: North
Holland).
\\ 12. G.T. Horwitz and A. Strominger, \NPB 360 1991 197,
S.S. Gubser, I.R. Klebanov and A.W. Peet, {\sl Preprint} {\bf PUPT-- 1598}
February 1996, {\bf hep-th/9602135}

}
}

\end{document}